\documentclass[iop12pt]{iopart}
\usepackage{setstack}
\usepackage{graphicx}
\usepackage{color}

\begin{document}

\title{Quantum simulation of neutrino oscillations with trapped ions}

\author{C Noh$^{1}$, B M Rodr\'{\i}guez-Lara$^1$ and D G Angelakis$^{1,2}$}
\address{$^1$ Centre for Quantum Technologies, National University of Singapore, 2 Science Drive 3, Singapore 117542.}
\address{$^2$ Science Department, Technical University of Crete, Chania, Crete, Greece, 73100}
\ead{cqtncs@nus.edu.sg, dimitris.angelakis@googlemail.com}
%\author{B. M. Rodr\'{\i}guez-Lara}
%\affiliation{Centre for Quantum Technologies, National University of Singapore, 2 Science Drive 3, Singapore 117542.}

%%\author{Dimitris G. Angelakis}
%\affiliation{Science Department, Technical University of Crete, Chania, Crete, Greece, 73100}
%\affiliation{Centre for Quantum Technologies, National University of Singapore, 2 Science Drive 3, Singapore 117542.}

\begin{abstract}
We propose a scheme to simulate the dynamics of neutrino oscillations using trapped ions. For neutrinos in 1+1 dimensions, our scheme is experimentally implementable with existing trapped ion technology. We show that the three generation neutrino oscillations can be realized with three ions for 1+3 and 1+1 dimensions where the latter case only requires experimentally proven two-ion interactions. For this case, we discuss two setups utilizing different types of spin-spin interactions. Our method can be readily applied to two generation neutrino oscillations requiring fewer ions and lasers. We give a brief outline of a possible experimental scenario.
\end{abstract}

\maketitle
\tableofcontents
\section{Introduction}
%\addcontentsline{toc}{section}{Introduction}
Ever since Pauli inferred their existence, the exact nature of neutrinos have been a mystery due to their tendency to avoid interacting with other particles. One prominent question was whether the neutrinos are massive particles. In 1957 Pontecorvo suggested that massive neutrinos can oscillate (change its flavour) and later noticed that it could explain the solar neutrino problem \cite{Pontecorvo}. Since then large scale experiments have confirmed that neutrinos have small but nonzero masses and give rise to oscillations between flavour eigenstates \cite{experiments}. The discovery of neutrino oscillations is regarded as one of the most important discoveries of modern elementary particle physics and have lead to many proposals on physics beyond the standard model. 

Recently there has been a growing interest in simulating exotic relativistic phenomena using other controllable quantum systems. There have been proposals to simulate equation in curved spacetime \cite{Boada}, Unruh effect \cite{Alsing}, and black-hole properties in BECs \cite{Garay}.  More recently, a scheme to simulate the many-body dynamics of a Dirac particle, viz., Schwinger effect, has also been proposed \cite{Szpak}. The seminal work on quantum simulation of the Dirac equation using trapped ions by Lamata \textit{et al.~}\cite{Dirac} and its subsequent experimental realization by Gerritsma \textit{et al.~}\cite{DiracE} are of particular interest to the current work. More recently, there has been a proposal \cite{Klein} and experimental demonstration \cite{KleinE} to simulate the Klein paradox.

On the other hand, trapped ion system is also one of the leading candidates for simulation of quantum spin systems. The idea was originally proposed by Porras and Cirac \cite{Porras} and recently similar schemes were verified experimentally. These experiments include the simulation of a quantum Ising model with two \cite{Friedenauer} or more \cite{Islam} ions, where phonon-mediated spin-spin interactions are realized \cite{Cirac, Milburn, Sorensen, Kim, Lee, Roos}. Combining the interactions giving rise to the linear momentum term used in the simulation of the Dirac equation and the spin-spin interactions used in the simulation of the Ising models simultaneously broadens the types of systems that can be simulated with trapped ions. For example, Casanova \textit{et al.~}have considered quantum simulation of the Majorana equation and unphysical operations \cite{Casanova}.

In this work, we show that neutrino oscillations can be simulated in trapped ion systems. Especially, observation of neutrino oscillations in 1+1 dimensions requires only the trapped ion technology demonstrated in experiments.  Because whether the neutrinos are Dirac or Majorana particles bear no observable consequences in neutrino oscillations, we assume, without loss of generality, that neutrinos are Dirac particles as described by the minimally extended standard model. In this model, the charged current interaction creates a flavour eigenstate which is in a superposition of mass eigenstates. Our scheme paves the way for an experimental study of neutrino oscillations with controllable creation of the initial state and oscillation length, allowing different types of neutrino oscillations experiments  (see, for example, \cite{textbook} about different types of neutrino oscillations experiments) in a single setup. Also, non-trivial initial states not observed in nature can be readily created. For example, a state in a superposition of positive and negative energy eigenstates produces Zitterbewegung-like high frequency oscillations \cite{Bernardini}.

\section{Standard theory of neutrino oscillations}
Here, we reproduce the gist of the standard theory of neutrino oscillations where the neutrinos are assumed to be created in a momentum eigenstate \cite{textbook}.
So far, experiments have verified that there are three flavours of neutrinos: electron, muon, and tauon neutrinos, which we denote as $\nu_\alpha$ with $\alpha = e,\mu ,\tau$. These flavor states are not mass eigenstates and therefore do not follow the dynamics given by the Dirac equation. They, however, are related to the mass eigenstates $\nu_k$ by a mixing matrix through the equation $|\nu_\alpha\rangle = \sum_{k} U^*_{\alpha k}|\nu_k\rangle$, where $k=1,2,3$ is used to label different mass eigenstates. Since a massive neutrino state obeys equation, one can write $\vert \nu_k(t)\rangle = e^{-iE_kt}\vert \nu_k\rangle$, where $E_k = \pm\sqrt{(c\vert \mathbf{p} \vert)^2 +(m_kc^2)^2}$, given that the state is in the momentum eigenstate with momentum $\mathbf{p}$. The time evolution of a flavor eigenstate is given by
\begin{eqnarray}
\vert \nu_\alpha (t) \rangle &= \sum_k U_{\alpha k}^{*} e^{-iE_kt}\vert \nu_k\rangle, \nonumber \\
& = \sum_\beta \sum_k U_{\alpha k}^{*} e^{-iE_kt} U_{\beta k} \vert \nu_\beta \rangle,
\end{eqnarray}
which means that the probability for the flavour of the neutrino to change from $\alpha$ to $\beta$ is
\begin{eqnarray}
P_{\nu_\alpha \rightarrow \nu_\beta}(t) = \sum_{k,j} U_{\alpha k}^{*} U_{\beta k}U_{\alpha j} U_{\beta j}^{*}e^{-i(E_k-E_j)t}.
\end{eqnarray}
In the ultrarelativistic limit $c|\mathbf{p}| \gg mc^2$
\begin{eqnarray}
E_k-E_j = \frac{\Delta m^2_{kj}c^4}{2E},
\end{eqnarray}
with $\Delta m^2_{kj} \equiv m_k^2 - m_j^2$ and $E\equiv c\vert\mathbf{p}\vert$. Thus, after replacing the propagation time $t$ with the distance traveled $L/c$, the probability becomes
\begin{eqnarray}
\label{eq5}
P_{\nu_\alpha \rightarrow \nu_\beta}(t) = \sum_{k,j} U_{\alpha k}^{*} U_{\beta k}U_{\alpha j} U_{\beta j}^{*}e^{-i\frac{\Delta m^2_{kj}c^3}{2E}L}.
\end{eqnarray}
Measuring the probability for flavour changes thus allows one to gain information about the squared mass difference and the mixing matrix. Especially, if all the masses are equal there would be no neutrino oscillations

\section{Trapped ion implementation}
Here, we show how the three generation neutrino oscillations in 1+1 D can be implemented in a system of 3 ions utilizing only previously tested ion manipulations. A similar implementation, extending the original scheme in \cite{Dirac} for 3+1 D is also possible, but as this scheme requires a slightly more complicated experimental setup not realized in the lab thus far, we focus on the 1+1 D case. The 3+1 D case is briefly discussed later.
\subsection{1+1 dimensions}
In 1+1 D the Dirac equation reads \cite{Thaller}
\begin{eqnarray}
\label{diracH}
i\frac{\partial\psi}{\partial t} = \left(c\hat{p}\sigma_x + mc^2\sigma_z \right)\psi,
\end{eqnarray}
which can be simulated by a single trapped ion with two internal levels \cite{Dirac, DiracE}. The ion is driven by a bichromatic laser that couples the internal levels with motional sidebands to create the Dirac Hamiltonian
\begin{eqnarray}
\label{DiracH}
H_D = 2\eta\Delta\tilde{\Omega}\sigma_x\hat{p} + \Omega\sigma_z,
\end{eqnarray}
where $\Delta = \sqrt{1/2\tilde{m}\omega}$ is the size of the ground state wave function with ion mass $\tilde{m}$; $\omega$ is the frequency of the vibrational mode coupled to the internal states via the bichromatic laser. $\eta$ is the Lamb-Dicke parameter, $\hat{p}$ is the momentum operator for the phonon mode, and the $\Omega$ term arises from the detuning $2\Omega$ between the bichromatic light field and the qubit transition. This Hamiltonian is equivalent to the Dirac Hamiltonian in 1+1 D with $c = 2\eta\tilde{\Omega}\Delta$ and $mc^2 =\Omega$. 

Neutrino oscillations arise from an interference between different energy eigenstates. To mimic this effect we need the correct relativistic energy dependence and an ability to create a superposition of different energy eigenstates, i.e.~for $\psi$ to be in a superposition of different mass eigenstates. For 1+1 D, each neutrino is described by a two-component spinor related to positive and negative energy states, which means that 6 basis states are required for 3 generations. We construct these basis states as follows: 
\begin{eqnarray}
\left| \nu_1 \right\rangle =  \left( \begin{array}{c} \alpha|ggg\rangle \\ \beta|geg\rangle \end{array} \right), 
\left| \nu_2 \right\rangle = \left( \begin{array}{c} \alpha|gge\rangle \\ \beta|gee\rangle \end{array} \right), 
\left| \nu_3 \right\rangle = \left( \begin{array}{c} \alpha|egg\rangle \\ \beta|eeg\rangle \end{array} \right),
\end{eqnarray}
where $|g\rangle$ and $|e\rangle$ denote the two internal states of a qubit.
It is easily seen that 
$1\otimes\sigma_x\otimes 1\left|\nu_k\right\rangle = \sigma_x\left|\nu_k\right\rangle$,
where the $\sigma_x$ on the r.h.s. exchanges the two basis states that define $\left|\nu_k\right\rangle$. Also, it is easy to work out that
\begin{eqnarray}
H_{ss}\left|\nu_1\right\rangle &= \left(\Omega_1 + \Omega_2\right)\sigma_z \left|\nu_1\right\rangle,\nonumber \\
H_{ss}\left|\nu_2\right\rangle &= \left(\Omega_1 - \Omega_2\right)\sigma_z \left|\nu_2\right\rangle, \nonumber \\
H_{ss}\left|\nu_3\right\rangle &= \left(-\Omega_1 + \Omega_2\right)\sigma_z \left|\nu_3\right\rangle,
\end{eqnarray}
where
$H_{ss} = \Omega_1 \sigma_z\otimes\sigma_z\otimes 1+ \Omega_2 1\otimes\sigma_z\otimes\sigma_z$.
Therefore with Eq.~(\ref{DiracH}) we see that the Hamiltonian
\begin{eqnarray}
\label{Ham1}
H =& 1\otimes H_D \otimes 1 + H_{ss}, \nonumber \\
=& 2\eta\Delta\tilde{\Omega}\left( 1\otimes\sigma_x\otimes 1\right)\hat{p} - \Omega 1\otimes\sigma_z\otimes 1 \nonumber \\ &+ \Omega_1 \sigma_z\otimes\sigma_z\otimes 1+ \Omega_2 1\otimes\sigma_z\otimes\sigma_z.
\end{eqnarray}
produces the correct dynamics for three generations of neutrinos. That is, the mass eigenstates follow the Hamiltonian
\begin{eqnarray}
H|\nu_k\rangle = \left( c\sigma_x\hat{p} +m_kc^2\sigma_z \right)|\nu_k\rangle, 
\end{eqnarray}
with neutrino masses
$m_1c^2 = \Omega +\Omega_1 + \Omega_2 $, 
$m_2c^2 = \Omega +\Omega_1 - \Omega_2 $, 
$m_3c^2 = \Omega -\Omega_1 + \Omega_2 $.
 
As mentioned earlier, the $H_D$ term can be created by focusing a detuned bichromatic laser on the second ion. The remaining part requires two-qubit gate type interactions on the qubits $(1,2)$ and $(2,3)$, respectively. One can get each of these two by selectively shining a pair of ions with a laser tuned to a particular phonon mode to mediate the ion-ion interaction \cite{Cirac, Milburn, Sorensen, Leibfried}. The two sets of lasers should act on two different normal modes to avoid interfering with each other. For example, the two gate type interactions can utilize the center of mass mode and the zigzag mode in a transverse direction, while the linear momentum part utilizes the axial mode. Then, for the Hamiltonian to work for all times, one needs $\eta \ll 1$ to avoid exciting a significant number of phonons which would result in qubit-phonon entanglement. Otherwise one could consider this as a gate operation which works only at certain times, in which case it is important to make sure that the two two-qubit interaction terms have commensurate gate times.

Recently, Kim \textit{et al.~}have managed to create a tunable spin-spin couplings between trapped ions without excitation of real phonons \cite{Islam, Kim}. This scheme offers nearly ideal spin-spin interactions that can be used to simulate neutrino oscillations. In their scheme all the ions are addressed simultaneously with two bichromatic laser beams whose optical beatnote detuning is far from each normal mode compared to that mode's sideband Rabi frequency. Thus the phonons are only virtually excited and the qubit states evolve according to the Hamiltonian
\begin{eqnarray}
\label{Ham2}
H'_{ss} = J_1\left( \sigma_x\otimes\sigma_x\otimes 1 + 1\otimes\sigma_x\otimes\sigma_x\right) + J_2\sigma_x\otimes 1\otimes\sigma_x,
\end{eqnarray} 
at all times. Furthermore, the signs and magnitudes of $J_1$ and $J_2$ can be controlled by changing the beatnote detuning and the spin-flip Rabi frequencies, which in turn allows one to control the effective neutrino masses. To make use of this Hamiltonian, the basis states $|g\rangle, |e\rangle$ should be changed to the x-basis states which we denote as $|0\rangle, |1\rangle$, and ions 1 and 2 should be swapped, i.e.
\begin{eqnarray}
\left| \nu_1 \right\rangle = \left( \begin{array}{c} \alpha|000\rangle \\ \beta|100\rangle \end{array} \right), 
\left| \nu_2 \right\rangle = \left( \begin{array}{c} \alpha|001\rangle \\ \beta|101\rangle \end{array} \right), 
\left| \nu_3 \right\rangle = \left( \begin{array}{c} \alpha|010\rangle \\ \beta|110\rangle \end{array} \right).
\end{eqnarray}
To mimic the dynamics of a neutrino, we add the linear momentum term and three extra single qubit lasers:
\begin{eqnarray}
\label{fullH}
H =& 2\eta\Delta\tilde{\Omega}\left( 1\otimes\sigma_y\otimes 1\right)\hat{p}  \nonumber \\ 
&+ J_1\left( \sigma_x\otimes\sigma_x\otimes 1 + 1\otimes\sigma_x\otimes\sigma_x\right) + J_2\sigma_x\otimes 1\otimes\sigma_x \nonumber \\
&+ J_1 1\otimes\sigma_x\otimes 1 + J_1 1\otimes 1\otimes\sigma_x -J\sigma_x\otimes 1\otimes 1,
\end{eqnarray}
yielding
\begin{eqnarray}
H|\nu_1\rangle = \left(2\eta\Delta\tilde{\Omega}\sigma_y\hat{p} + \sigma_z \left(J + J_1 + J_2 \right)\right)|\nu_1\rangle -J_1|\nu_1\rangle, \nonumber \\
H|\nu_2\rangle = \left(2\eta\Delta\tilde{\Omega}\sigma_y\hat{p} +\sigma_z\left(J + J_1 - J_2\right)\right)|\nu_2\rangle - J_1|\nu_2\rangle, \nonumber \\
H|\nu_3\rangle = \left(2\eta\Delta\tilde{\Omega}\sigma_y\hat{p} +\sigma_z\left(J -J_1 + J_2\right)\right)|\nu_3\rangle - J_1|\nu_3\rangle.
\end{eqnarray}
Ignoring the constant term $J_1$, we get the Hamiltonian that describes 3 types of neutrinos with $m_1 c^2 = J + J_1 + J_2$, $m_2 c^2 = J + J_1 - J_2$, and $m_3 c^2 = J - J_1 + J_2$.
Note that for the momentum term we now use an alternative but equivalent form $\sigma_y \hat{p}$, which can be implemented in the same way as the original scheme by changing the phase of the laser; also, we have assumed that the detuning is zero

%%%%%%%%%%%%%%%%%%%%%%%%%%%%%%%%%%%%%%%%%%%%%%%%%%%%%%%%%%%
\begin{figure}[ht]
%\psfrag{y}[][]{\footnotesize $g^{(2)}_{\uparrow \downarrow}(l), g_{-}^{(2)}(l)$}
\hspace{2.5cm}\includegraphics[width= 7.0cm]{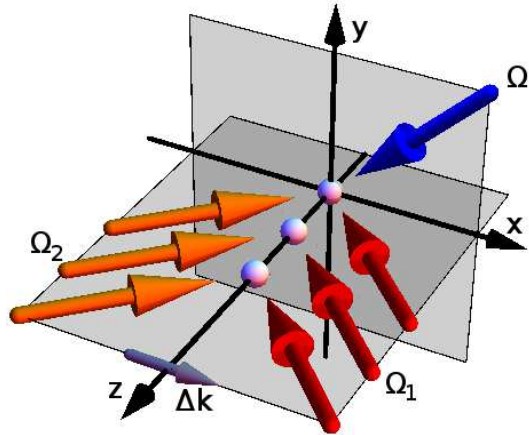}
\caption{A potential experimental setup to simulate neutrino oscillations. The lasers $\Omega_1$ and $\Omega_2$ shine on all 3 ions with a given beatnote frequency and a wavevector difference $\Delta k$ in a transverse direction, to create the mass terms with the help of an extra laser acting on each ion (not shown on the figure). $\Omega$, acting on the second ion, creates the linear momentum term of the Dirac Hamiltonian.}\label{fig:Fig1}
\end{figure}
%%%%%%%%%%%%%%%%%%%%%%%%%%%%%%%%%%%%%%%%%%%%%%%%%%%%%%%%%%%%%
The experimental setups of  Islam \textit{et al.~}\cite{Islam} and Kim \textit{et al.~}\cite{Kim} use transverse phonon-modes to mediate the spin-spin interactions which means that the momentum part could utilize an axial mode or the perpendicular transverse mode. Then, three extra lasers with appropriate intensities and phases help create the appropriate mass terms. Figure 1 shows a possible experimental setup to simulate neutrino oscillations; for visibility three extra lasers addressing each ion are not shown on the figure.

We also note that the rotated version of Eq.~(\ref{Ham1}) (so that the gate operations $\sigma_z\otimes\sigma_z$ are replaced by $\sigma_y\otimes\sigma_y$  for example) can be directly implemented using spatially dependent Rabi frequencies as opposed to a uniform Rabi frequency used in deriving the Hamiltonian in Eq.~(\ref{Ham2}) \cite{Kim2}.

For two generations, only two trapped ions are required, which is interesting not only because they have less parameters, but also because some experiments are not sensitive to all three generations and can be described by an effective model with two-neutrino mixing \cite{textbook}. The two generation case is significantly simpler than the three generation case as only one spin-spin coupling term and one single-qubit rotation term are needed to generate the mass terms.

So far we have proposed schemes that utilize multiple trapped qubits. However, if one could find a stable multi-level trapped ion system, it is possible to simulate neutrino oscillations. For example, if there are 3 ground states and 3 excited states and a single bichromatic laser addressing the three transitions, one has the Dirac Hamiltonian (\ref{DiracH}) for each transition. If the transitions are not of the same frequency the masses would be different by default, otherwise one could use external fields to shift the energy levels. Possible difficulties in such single-ion schemes are short decoherence time and preparation of a general initial state.

\subsection{3+1 dimensions} 
Here, we give a brief description of how to simulate the full 3+1 D dynamics using the 4-lv scheme introduced in \cite{Dirac}. We can get the Dirac Hamiltonian for 3 generations by considering the following basis states:
$\left| \psi_1 \right\rangle = (|a\rangle , |b\rangle , |c\rangle , |d\rangle ) \otimes |a\rangle\otimes |a\rangle$,
$\left| \psi_2 \right\rangle = ( |a\rangle , |b\rangle , |c\rangle , |d\rangle ) \otimes |b\rangle\otimes |a\rangle$, 
$\left| \psi_3 \right\rangle = (|a\rangle , |b\rangle ,  |c\rangle , |d\rangle )\otimes |a\rangle\otimes |b\rangle$. 
Apart from the Hamiltonian proposed in \cite{Dirac} acting on the first ion, we add the following spin-spin interaction terms
$-\Omega_1 (\sigma_y^{ac}-\sigma_y^{bd})\otimes\sigma_z^{ac}\otimes 1 - \Omega_2 (\sigma_y^{ac}-\sigma_y^{bd})\otimes\sigma_z^{bd}\otimes 1 \nonumber \\ - \Omega_3 (\sigma_y^{ac}-\sigma_y^{bd})\otimes 1 \otimes\sigma_z^{bd}$
to obtain the correct relativistic equations with $m_1c^2 =  (\Omega +\Omega_1 )$, $m_2c^2 =  (\Omega +\Omega_2 )$, and $m_3c^2 =  (\Omega +\Omega_1 +\Omega_3 )$,
as can be easily verified. In most of the realistic cases there is only a single non-zero momentum component which reduces the number of lasers needed. 

\section{A possible experimental scenario}
Once the Hamiltonian is engineered an appropriate initial state has to be prepared to simulate neutrino oscillations. For example, if the initial state is in a mass eigenstate one would observe no oscillations. These cases are however quite special and flavours would oscillate for a generic initial state. A physically interesting case is when the initial state is in a definite flavour state and has a momentum wave packet with a narrow momentum distribution around an average momentum in the ultrarelativistic regime.

We describe a scenario where an electron neutrino is created which propagates for a certain amount of time before it is detected. For concreteness we use the implementation that uses $\sigma_z\otimes\sigma_z$ type interactions.  An electron neutrino state can be written as
\begin{eqnarray}
|\nu_e(p)\rangle = \frac{1}{\sqrt{3}}\left( \sqrt{2}|\nu_1(p)\rangle - |\nu_2(p)\rangle\right),
\end{eqnarray}
according to a mixing matrix called tribimaximal mixing matrix that is consistent with experiments \cite{Harrison, Nunokawa}. Note that this state only involves 2 states and thus can be described by a two generation model, i.e. with two ions. Since this state is not entangled:
\begin{eqnarray}
|\nu_e(p)\rangle 
= |g\rangle\otimes\left( \alpha_p|g\rangle + \beta_p|e\rangle\right)\otimes\frac{1}{3}\left(\sqrt{2}|g\rangle - |e\rangle \right),
\end{eqnarray}
it can be prepared by performing single-qubit rotations on the ground state $|ggg\rangle$.
Other flavor states are entangled and need 2 qubit gate operations. However, it has been shown that arbitrary states can be created by repeated use of gate operations \cite{Lloyd}, and one such algorithm for trapped ions has already been proposed \cite{Sasura}. A similar proposal to prepare hadronic states in terms of up and down quarks and their spins has been given recently \cite{Semiao}. The next step is to engineer the state of the phonon mode (momentum distribution). To create a state with a given average momentum, one would cool down the ions to the ground state and apply a momentum-displacement operation \cite{DiracE}. An arbitrary state can in principle be constructed by performing state-dependent displacement operations; positive or negative energy eigenstates have asymmetric spinor components that depend on the momentum of the particle and a superposition of momentum eigenstates gives a physical wavepacket of the particle. However, it is possible to approximate an energy eigenstate with a Gaussian momentum distribution for a Dirac particle in 1+1 D by approximating the eigenstate with a spinor that has an asymmetry between the average momenta of the components \cite{DiracE}.  These approximate states can be created by focusing a momentum displacement laser on the second ion. Note that the asymmetry in the momentum distribution between the spinor components becomes smaller as one increases the average momentum of the particle, so in the ultrarelativistic regime the momentum wave function approaches the symmetric spinor $\psi(p) \propto \mathrm{exp}[-(p-p_0)^2/2\sigma](1,1)$ with average momentum $p_0$.
%%%%%%%%%%%%%%%%%%%%%%%%
\begin{figure}[h]
%\psfrag{y}[][]{\footnotesize $g^{(2)}_{\uparrow \downarrow}(l), g_{-}^{(2)}(l)$}
\hspace{2.5cm}\includegraphics[width= 8.0cm]{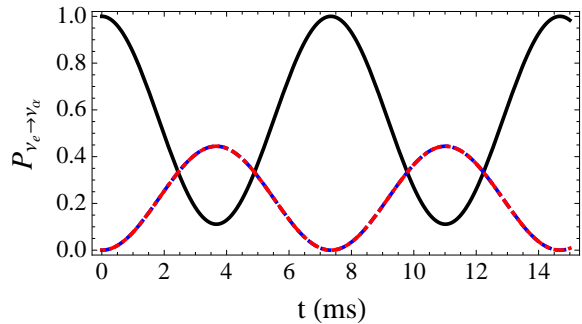}
\caption{Neutrino oscillations of an electron neutrino in a momentum eigenstate. The top (black) curve represents electron component whereas the bottom (blue and red, overlapping) curves represent muon and tauon neutrino components. The kinetic energy is $2\pi\times$40 Khz and the rest mass energies are $2\pi\times$(5, 6, and 7) Khz.}\label{fig:Fig2}
\end{figure}
%%%%%%%%%%%%%%%%%%%%%%%%%%%
After creating the initial conditions the interactions can be switched on and the required states would be observed after waiting enough time for a significant flavour change to have occurred. The flavour change means changing probability amplitudes for the mass eigenstates which corresponds to changing internal states. In principle, the full internal state can be measured using quantum state tomography \cite{Roos2} after tracing out the phonon modes. However, the different mass eigenstates can be made to have different fluorescence rates (with extra single qubit rotations) when coupled to an auxiliary level via an external laser field and therefore be distinguished by looking at the fluorescence level \cite{Friedenauer}. Then, the measured fluorescence level would oscillate in accordance with the neutrino flavour components. Exactly how it oscillates would depend on a particular implementation used, but it can be calculated theoretically and then compared to experimental results. Figure \ref{fig:Fig2} shows, as an example, flavour oscillations of an electron neutrino created at $t=0$, calculated from Eq.~(\ref{eq5}) with the tribimaximal mixing matrix. The kinetic and rest mass energies, shown in the caption, are chosen to correspond to experimentally viable numbers while obeying the ultrarelativistic condition. The oscillations should clearly be visible in real experiments with decoherence, as the decoherence time can be of the order of 10ms (see e.g. \cite{Kim2}). The oscillation period decreases if one moves closer to the normal realtivistic regime where the kinetic and the mass energies are similar, easing the requirement for a required decoherence time.

\section{Conclusion}
We have proposed an experimentally feasible scheme to simulate two or three generation neutrino oscillations using trapped ions. In 1+1 dimensions, our proposal only utilizes experimentally proven techniques, allowing a controlled experimental observation of neutrino oscillations. In this work, due to its relevance to neutrino oscillations experiments, we have assumed that an initial electron neutrino is in an energy eigenstate with a momentum distribution. However, other initial conditions can also produce neutrino oscillations and could provide interesting alternative scenarios, e.g. outside the ultrarelativistic regime or initial states that are not energy eigenstates. Also, the current setup allows simulations of different types of neutrino oscillations experiments in a single experimental setup.

\section*{Acknowledgements}
The authors would like to acknowledge the financial support by the National Research Foundation and Ministry of Education, Singapore.

\section*{References}
\bibliographystyle{unsrt}

\end{document}